\documentclass[smallextended]{svjour3}       
\usepackage{graphicx}
\begin{document}

\title{Momentum and energy propagation in tapered granular chains
}


\author{L. P. Machado        \and
        Alexandre Rosas \and Katja Lindenberg
}


\institute{L. P. Machado \at
              Departamento de F\'isica, CCEN, Universidade Federal da Para\'iba, Caixa Postal 5008, 58059-900, Jo\~ao Pessoa, Brazil
           \and
           Alexandre Rosas \at
Departamento de F\'isica, CCEN, Universidade Federal da Para\'iba, Caixa Postal 5008, 58059-900, Jo\~ao Pessoa, Brazil\\
	\email{arosas@fisica.ufpb.br}
	\and
	Katja Lindenberg \at
Department of Chemistry and Biochemistry and BioCircuits Institute, University of California San Diego, La Jolla, CA 92093-0340, USA
}

\date{Received: date / Accepted: date}

\maketitle

\begin{abstract}
We study momentum and energy propagation in 1D tapered 
chains of spherical granules which interact according to a Hertz 
potential. In this work we apply the binary collision approximation, 
which is based on the assumption that transfer of energy along the 
chain occurs via a succession of two-particle collisions. Although the 
binary theory correctly captures the trends of increase or decrease of 
kinetic energy and momentum, the actual values of these quantities are 
not in good quantitative agreement with those obtained by numerically
integrating the full equations of motion.  To address this difficulty we have
developed a mixed numerical/analytical correction algorithm to provide an improved
estimate of the velocity of the particles during pulse propagation.
With this corrected velocity we are in turn
able to correctly predict the momentum and kinetic energy along 
the chain for several tapering configurations, specifically for forward linear, 
forward exponential, backward linear and backward exponential tapering.

\end{abstract}

\section{Introduction}
\label{intro}

Pulse propagation through granular media has become a topic of 
intense theoretical, numerical, and experimental attention because 
it involves fundamental issues in nonlinear discrete systems, and 
because at the same time it is of major practical
importance. When a granular chain of equal spherical granules without precompression and without
gaps is hit at one end, a solitary wave travels down the chain essentially unchanged in time.  This
observation was first made by Nesterenko~\cite{nesterenko83,nesterenko94,book-nesterenko}, who solved this
nonlinear problem using an analytic continuum approximation and successfully compared his analytic solutions to numerical simulation results.  Experimental verification followed
soon after ~\cite{nesterenko85} and later in~\cite{falcon97}.  The way in which a pulse propagates along a chain is strongly dependent
on the characteristics of the granules~\cite{hinch99}. By modifying such a chain in various ways, e.g., by
tapering, including large and small masses, adding impurities to the chain, and other such
manipulations, it was soon realized that it is possible to control exactly how the pulse energy
behaves as it propagates~\cite{sen01,wu02,nakagawa,doney05,sen05b,sen06,melo06,doney09,daraio,job05,hascoet00,hascoet99}
or becomes trapped with slow leakage~\cite{hong,wang07}. This control offers the possibility of
designing granular media with carefully chosen shock absorption or signal propagation properties.
Other interesting configurations leading
to solitary wave train formation were explored~\cite{sen07,melo07}.
The feasibility of optimizing granular configurations for clean reflection of a pulse from a wall
was studied as a tool for nonintrusive or nondestructive sensors~\cite{job05}.  These and many
other applications were discussed and reviewed in~\cite{SEN,sen08}.  The effects of various forms of
friction on pulse propagation were analyzed in~\cite{alexandrefriction,alexandre07,alexandre08},
including the observation of entirely new waveforms as a result of momentum-conserving
frictional forces.  On the theoretical front, a quantitatively accurate theory was first developed
three decades ago for chains made up of identical spherical granules with no precompression
and no gaps~\cite{nesterenko83,nesterenko94}.  This approach was difficult to generalize to other granular
configurations.  As an alternative, we developed a theory based on a binary collision approximation (BCA)
which has been generalized to tapered, and decorated chains in various initial states
and even to chains with some random configurational
elements~\cite{alexandrefriction,alexandre04,our-tapered,our-decorated,decorated2010,italo}. This
theory has been quantitatively accurate in many cases, and forms the basis for the discussion in
this paper.

Experiments on granular chains has been restricted 
by the constraints of particular experimental setups.
Analytic work can overcome such strictures and is helpful in expanding our
fundamental understanding of these systems. The methods we have 
recently developed~\cite{alexandre04,our-tapered,our-decorated,decorated2010,italo}
have proved broadly 
applicable to chains of a variety of configurations. Our analytic BCA
works remarkably well not only qualitatively
but also quantitatively.
Even though the BCA 
is restricted to conservative chains (elastic collisions), and real granular collisions are not
entirely elastic, the BCA is nevertheless able to capture experimentally observed
dynamics of chains of very hard spheres moving along very highly polished
channels. This is the case with the experiments in the references cited above. 

Why would one want to study chains of these particular configurations 
(such as tapered or decorated)? In addition to questions of theoretical 
interest, the reason is that such physical arrangements of granules 
provide the opportunity to optimize the shock absorption properties of
 the system, and it is hoped that what one learns in one dimension can 
 guide the inquiries in more realistic higher dimensional systems. It 
 is quite surprising how exquisitely sensitive the transmittal of 
 energy and/or momentum along a chain can be to details of chain
granule configuration. The characterization of energy
and momentum propagation along a granular chain is in itself quite 
varied.  For instance, one might ask how long it takes the maximum of 
a pulse to reach a given granule.  Or one might wish to know how long 
the pulse maximum resides on a particular granule, or how much energy 
is focused on that granule during this time of residence. One might 
wish to know what happens to the width of a pulse as it travels along 
a chain, whether or not the pulse retains its integrity, and one might 
wish to predict whether or not the chain will fragment in the
process. Particular quantities of interest in the assessment of shock 
absorption properties of these simplified one-dimensional granular
systems are the momentum and energy with which a pulse that starts at 
one end of a granular chain would first hit a wall at the other end.  
In our earlier work we focused on the pulse velocity amplitude and on 
 the time that the pulse spends on each granule before moving on to the 
 next granule.

The BCA assumes that the traveling pulse resides on only two granules
at a time, so that the many-body problem of a fully interacting chain
is reduced to an iterated solution of successive two-body problems. 
The approximation is straightforwardly implemented for
monodisperse~\cite{alexandre04} and tapered 
chains~\cite{our-tapered}. For decorated
chains, that is, chains of large (possibly tapered)
grains separated by one or more small granules, it is first necessary to 
replace the chain by an effective 
monodisperse or tapered one to which the approximation can then be 
applied~\cite{our-decorated,decorated2010}.
The approximation relies on the assumption that the
traveling pulse is very narrow, which holds for many granular 
geometries (including spherical granules).  Within  the binary 
collision approximation, we were able to make exceptionally good 
predictions for certain pulse properties. However, not every prediction 
of the BCA is equally quantitatively accurate. In particular, even 
though the change of the velocity of the particles, $dv_k/dk$, as a 
pulse moves along a chain of granules labeled by $k$ is predicted 
extremely accurately, the BCA consistently over-estimates the actual
value $v_k$ of the velocity amplitude itself. The reason is that in 
reality the pulse does not reside precisely on only two granules 
 at a time because the interactions are not hard-sphere.
 
In this work we focus on correcting the error made in the calculation 
of the velocities of the granules during pulse propagation 
along a chain of initially uncompressed spherical grains that just 
touch each other, which in 
turn affects our predictions for momentum and energy propagation.  As 
we shall see, the corrections are mild but important. We introduce 
corrections that are partly analytical and partly numerical, and that
provide a set of tools for further quantitative prediction.

Before launching into our presentation we clarify a point that might otherwise appear
puzzling.  As the energy pulse travels down our various chains, we observe that the momentum
and energy appear not to be conserved.  The momentum may grow or decay, and the energy
may appear constant or in fact may also decay.  This occurs because we are specifically
focusing on the pulse that carries energy forward.  At each collision event, there may also
be a backscattering of granules that also carry energy and momentum, and it may also happen that some portion
of the energy may lag behind in the grains behind the pulse.  When this is taken
into account, all is as it should be with energy and momentum conservation.

We organize our presentation as follows. In Sec.~\ref{oldequations} we briefly recap the equations of motion that
form the basis of all the work to follow. 
In Sec.~\ref{back} we present the results of our previous
calculations for backward tapered chains and in Sec.~\ref{for} for forward tapered chains.
In Sec.~\ref{newresults} we discuss the corrections
that lead to more accurate predictions for energy and momentum propagation along tapered
chains.  Section~\ref{collection} is a summary of our results.

\section{Equations of motion}
\label{oldequations}

We consider a one-dimensional chain of spherical granules~\cite{note} of density $\rho$, labeled by index $k$, with radius $R_k'$ and mass $M_k=(4/3)\pi\rho(R_k')^3$. 
The equation of motion for the displacement $y_k$ of the $k$th granule as a function of time $\tau$ is
\begin{equation}
M_k\frac{d^2y_k}{d\tau^2} = a r_{k-1}'(y_{k-1}-y_k)^{3/2}\theta(y_{k-1}-y_k) - a r_k'(y_k-y_{k+1})^{3/2} \theta(y_k-y_{k+1}),
\label{motion1}
\end{equation}
with $r_k' = \left[2(R_k' R_{k+1}'/(R_k'+R_{k+1}')\right]^{1/2}$ and $a$ a constant determined by Young's modulus and Poisson's ratio.
The Heaviside function $\theta(y)$ ensures that the elastic interaction between grains only exists if they are in contact.  Initially the granules just touch their neighbors in their equilibrium positions, that is, there is no pre compression and there are no gaps, and all but the leftmost particle are at rest. The initial velocity of the leftmost particle, $k=1$, is $V_1$.  

It is convenient to define scaled variables and parameters as follows,
\begin{eqnarray}
x_k=\frac{y_k}{R_1' \alpha^{2/5}}, &\qquad& t=\frac{V_1 \tau}{R_1' \alpha^{2/5}},\\
R_k=\frac{R_k'}{R_1}, &\qquad& m_k=\frac{M_k}{M_1},
\end{eqnarray}
with $ \alpha = M_1 V_1^2 / a (R_1')^{n+1/2}.$

Equation~(\ref{motion1}) can then be rewritten in the cleaner form, which we use for all subsequent analysis,
\begin{equation}
m_k\frac{d^2x_k}{dt^2} =  r_{k-1}(x_{k-1}-x_k)^{3/2}\theta(x_{k-1}-x_k) -  r_k(x_k-x_{k+1})^{3/2} \theta(x_k-x_{k+1}),
\label{motion2}
\end{equation}
with $r_k = \left[2(R_kR_{k+1}/(R_k+R_{k+1})\right]^{1/2}$. The rescaled initial velocity is unity, $v_1(t=0)=1$, as is the mass $m_1 =1$ and the radius $R_1$ of the first granule.  The velocity of the $k$th granule in unscaled variables is simply $V_1$ times its velocity in the scaled variables. The equations of motion can be integrated numerically. All of our previous results are obtained beginning from this starting point.

In a series of papers~\cite{alexandre04,our-tapered,our-decorated,decorated2010} we went on to solve these equations of motion using an approximation that we called the Binary Collision Approximation (BCA).  The approximation is based on the assumption that the transfer of energy along the chain occurs via a succession of two-particle collisions.  Particle $k=1$ of unit velocity collides with initially stationary particle $k=2$, which then acquires velocity $v_2$ and collides with initially stationary particle $k=3$, and so on. The velocities after each collision can easily be obtained from conservation of energy and momentum. 
This scheme is straightforward to implement if the chain is monodisperse or if it is tapered forward (decreasing radii) or backward (increasing radii).  If the chain is ``decorated" with small grains among large ones, then the decorated chain must first be replaced by a carefully constructed effective chain of the monodisperse or tapered variety to which the BCA can then be applied.  Here we exhibit the final full analytic BCA result, which requires the application of a particular tapering protocol for further implementation in later sections.  The velocity according to the BCA is 
\begin{equation}
  v_k=\prod_{k'=1}^{k-1} \frac{2}{1+\frac{\displaystyle m_{k'+1}}{\displaystyle m_{k'}} }.
\end{equation}  
Together with the mass $m_k=(R_k/R_1)^3$ this will then yield the pulse momentum $P_k=m_k v_k$ and the pulse energy $E_k=(1/2)m_k v_k^2$.  

As stated above, the BCA works very well for some quantities but is not quantitatively correct in its prediction of the pulse velocity.  This quantitative discrepancy between approximate and exact results thus leads to errors in the prediction of the momentum and the energy of the pulse; the energy dependence on the square of the velocity of course magnifies the problem.  We thus searched for a way to improve the theory.  We call this improvement a ``modified" theory.  In the next two sections we simply present, without re-deriving, the results of the original BCA (in scaled variables) relevant to this discussion.  The modified theory is then presented in Sec.~\ref{newresults}.

\section{Backward tapered chains}
\label{back}

We start with a collection of the results obtained for backward tapered chains using the BCA.
Backward tapered chains are alignments of granules of progressively increasing size
and/or mass. We assume that the granules are all made of the same material, so that they
have the same density and elastic properties. We study two different rules for the
way that the radii $R_k$ increase along the chain, whose granules are labeled by index $k$.
In linearly tapered chains, the radii increase arithmetically, 
\begin{equation}
  R^{bl}(k) = 1 + S \left( k-1 \right),
  \label{eq:radii-linear-backward}
\end{equation}
while exponential tapering is one in which the radii increase geometrically,
\begin{equation}
  R^{be}(k)= \frac{1}{(1-q)^{k-1}}.
  \label{eq:radii-exp-backward}
\end{equation}
$S>0$ and $0<q<1$ are the tapering parameters. Here, the radii of the particles are given in units 
of the radius of the leftmost granule (the impacting granule). The subscripts stand for ``backward linear" and ``backward exponential" respectively.

\subsection{Linearly tapered chains}

\begin{figure}[ht]
\centering
\includegraphics[width=75mm,angle=0]{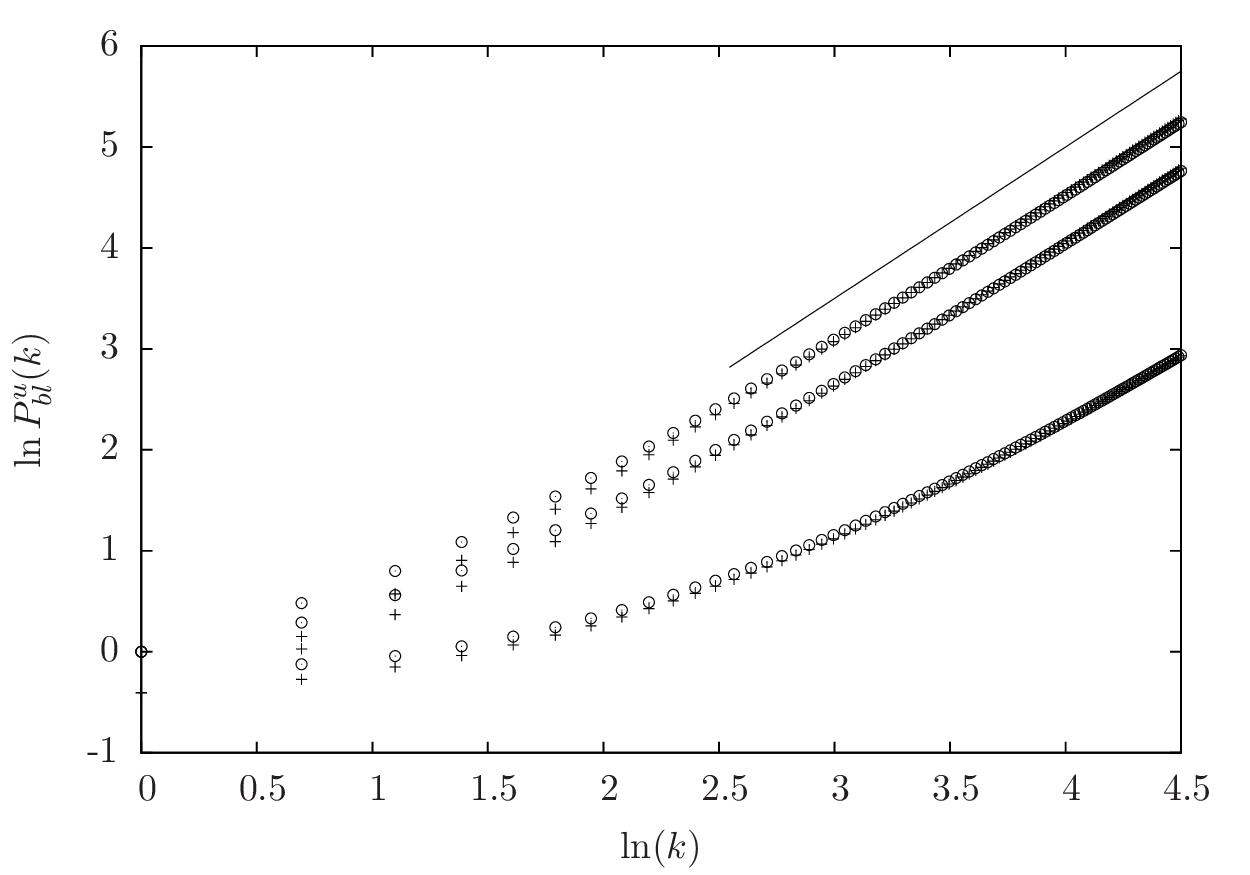}
\caption{Momentum as a function of granule number $k$ for backward
linearly tapered chains for
$S = 0.1$ to $0.9$ from bottom to top, in steps of $0.4$. The circles correspond to the numerical data, the + symbols represent the BCA and the solid line corresponds
to an arbitrary line with the slope 3/2 predicted by the BCA. 
\label{fig:p-k-logscale-lin}}
\end{figure}

\begin{figure}[ht]
\centering
\includegraphics[width=75mm,angle=0]{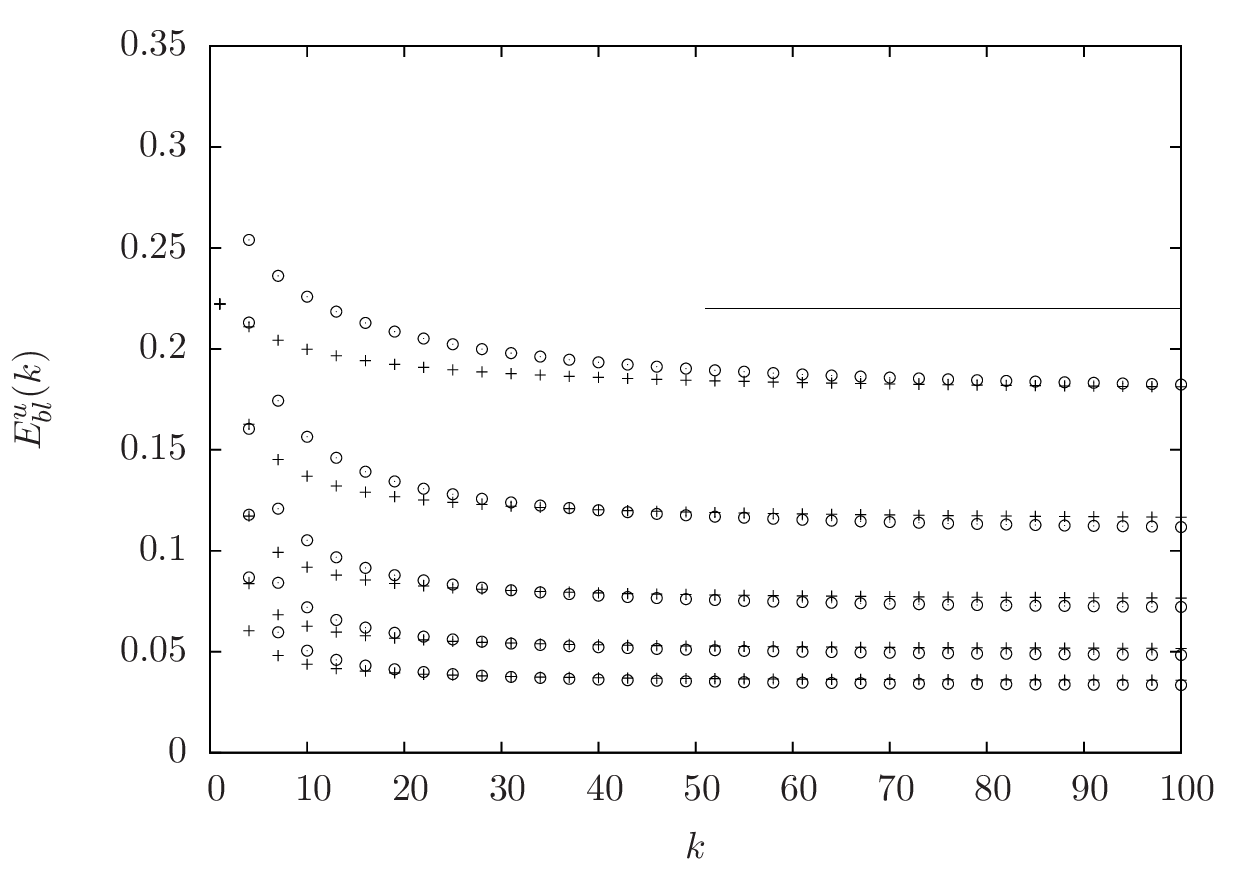}
\caption{Kinetic energy as a function of $k$ for backward
linearly tapered chains for
$S = 0.1$ to $0.9$ from top to bottom, in steps of $0.2$. The circles correspond to the numerical data, the + symbols represent the BCA and the solid line is an
arbitrary horizontal line indicating the asymptotic behavior predicted by the BCA.
\label{fig:E-k-asympt-lin}}
\end{figure}

Using energy and momentum conservation, one can show that for linearly tapered chains
the maximum velocity $v_{bl}^u(k)$ of granule $k$ as the pulse travels along the
chain varies as~\cite{our-tapered}
\begin{equation}
v_{bl}^{u}(k) \sim k^{-3/2}
\label{vltc}
\end{equation}
for large $k$.  The superscript indicates that these are results of the original \emph{unmodified} BCA theory.
On the other hand, since $m\sim R^3$ for any configuration, 
\begin{equation}
m_{bl}(k) \sim k^3.
\end{equation}
The masses of the granules are given in units of the mass of the
leftmost particle.
Therefore, the momentum and kinetic energy are predicted to have the asymptotic behavior
\begin{eqnarray}
  P_{bl}^{u}(k)&=& m_{bl}(k) v_{bl}^{u}(k) \sim k^{3/2}  \label{eq:momentum-lin},\\
  E_{bl}^{u}(k) &=& \frac{1}{2} m_{bl}(k) \left[v_{bl}^{u}(k)\right]^2 \sim 1.
  \label{eq:energy-lin}
\end{eqnarray}
It is worth mentioning that the exact form of $v_{bl}^{u}$, and
consequently of $E_{bl}^{u}$ and $P_{bl}^{u}$, can be obtained as a closed 
product expression if we use the expressions given in Sec.~\ref{oldequations}. However, for purposes
of discussion it is here
more instructive to focus on the asymptotic behaviour as given in Eqs.~(\ref{vltc}), 
(\ref{eq:momentum-lin}) and (\ref{eq:energy-lin}). We stress that all the BCA data 
presented in this manuscript was obtained using the full analytic
predictions of the theory and not just the asymptotic results. 
In Fig.~\ref{fig:p-k-logscale-lin}, we show that for various values of the tapering
parameter $S$ the momentum growth follows the trend predicted by
Eq.~(\ref{eq:momentum-lin}) -- compare the slopes of the different sets of symbols with
the slopes of the lines.
The energy, in turn, exhibits the saturation behavior predicted by
Eq.~(\ref{eq:energy-lin}), as can be seen from Fig.~\ref{fig:E-k-asympt-lin}.

\subsection{Exponentially tapered chains} 

\begin{figure}[ht]
\centering
\includegraphics[width=95mm]{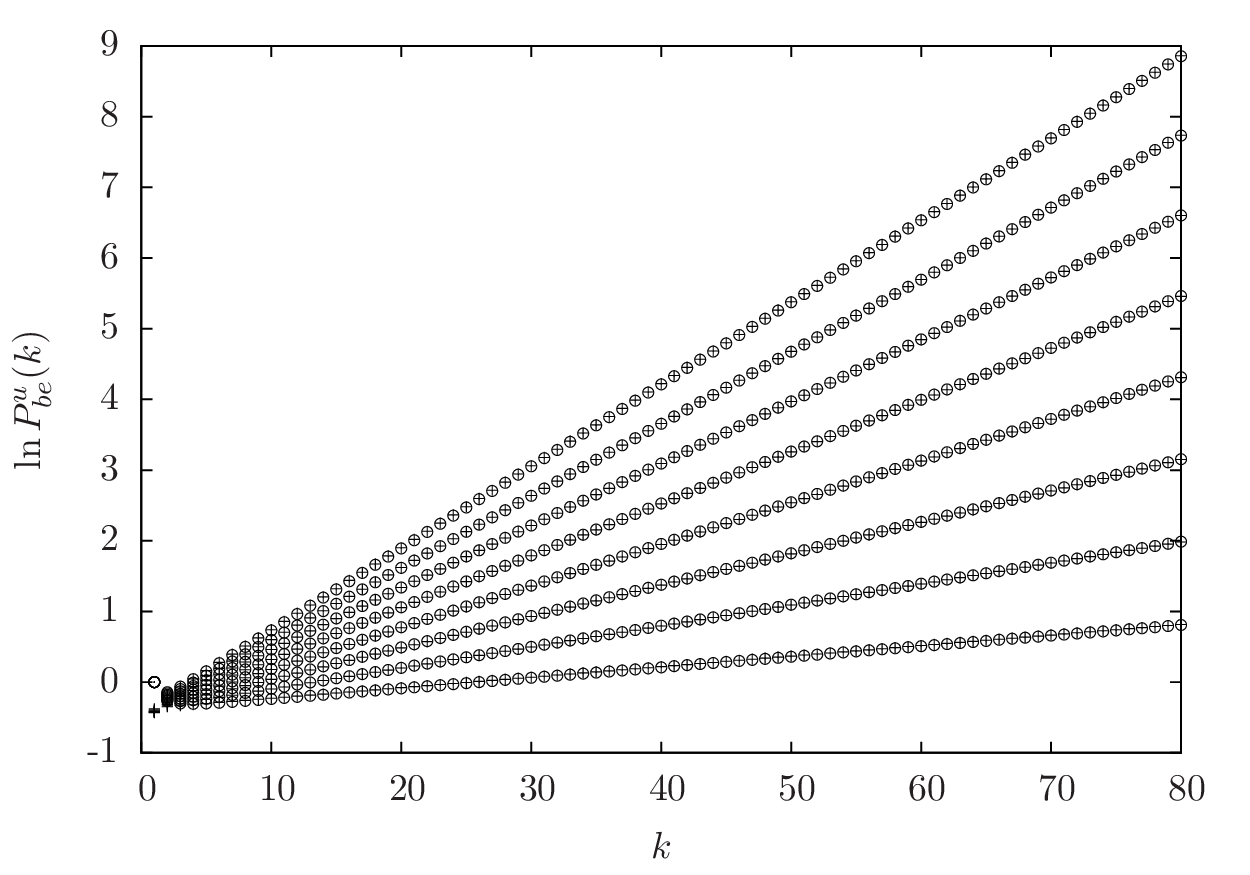}
\caption{Momentum as a function of $k$ for backward exponentially tapered
chains for $q = 0.01$ to $0.08$ from bottom to top, in steps of $0.01$. The circles correspond to the numerical data and the + symbols represent the BCA.\label{fig:p-k-logscale-exp}}
\end{figure}

For the backward exponentially tapered chain the velocity in the BCA and the
mass behavior are respectively
(see~\cite{our-tapered})
\begin{eqnarray}
\label{vetc}
  v_{be}^{u}(k) &=& A_{be}^u(q) e^{-k \ln A_{be}^u(q)}, \\
  m_{be}(k) &=& \left( 1 - q \right)^{3(1-k)},
  \label{eq:m-v-exp}
\end{eqnarray}
with
\begin{equation}
A_{be}^u(q) = \frac{1}{2} \left[ 1 + \left( 1-q \right)^{-3} \right]
\end{equation}
Hence, the momentum in the BCA for all $k$ (not just asymptotically) is given by
\begin{equation}
  P_{be}^u(k) = \left( 1 - \frac{3}{2} q + \frac{3}{2} q^2 - \frac{1}{2} q^3  \right)^{1-k}.
  \label{eq:momentum-exp-bin}
\end{equation}
This expression can be compared with the results of the numerical integration of
the equations of motion as follows. If we take the natural logarithm of
Eq.~(\ref{eq:momentum-exp-bin}), we see that the BCA prediction for the momentum
as a function of $k$ is an exponential growth characterized by 
\begin{equation}
{\mathcal P}_{be}^u \equiv \frac{\ln[P_{be}^u(k)]}{k}\sim  -\ln\left( 1 - \frac{3}{2} q +
\frac{3}{2} q^2 - \frac{1}{2} q^3 \right).
\label{rexpk}
\end{equation}
An exponential increase of the momentum can indeed be seen in Fig.~\ref{fig:p-k-logscale-exp}.
Furthermore, the rate of increase is in good agreement with the BCA
prediction. In Table~\ref{tab:coef-momentum}, the first two columns compare the exponential
rate of growth according to the BCA theory and the numerical results, respectively.
Exceptional agreement up to the second or third decimal digits confirm the quality of
the approximation.

\begin{table}[ht]
\centering
\begin{tabular}{|c|c|c|c|c|}
\hline
    & \multicolumn{2}{|c|}{Backward} & \multicolumn{2}{|c|}{Forward} \\ \hline
$q$ & BCA & Numerical & BCA & Numerical \\ \hline \hline
 0.01 &   0.01496 & 	0.01491 & -0.0150369 &	-0.0151371\\ \hline
 0.02 &   0.02985  & 	0.02972 & -0.030145  &	-0.0305032\\ \hline
 0.03 &   0.04465 & 	0.04440 & -0.0453208 &	-0.0462143\\ \hline
 0.04 &   0.05936 & 	0.05896 & -0.0605606 &	-0.0623547\\ \hline
 0.05 &   0.07398  & 	0.07340 & -0.0758609 &	-0.0790053\\ \hline
 0.06 &   0.08851 & 	0.08773 & -0.0912182 &	-0.0962407\\ \hline
 0.07 &   0.10294  & 	0.10194  & -- & -- \\ \hline
 0.08 &   0.11727  & 	0.11605  & -- & -- \\ \hline
\end{tabular}
\caption{Comparison of the increase as characterized by ${\mathcal P}_{be}^u$  and the decrease as characterized by 
${\mathcal P}_{fe}^u$  of the momentum as predicted by the BCA theory and the results
obtained from the numerical integration of the equations of motion.}
\label{tab:coef-momentum}
\end{table}

\begin{figure}[ht]
\centering
\includegraphics[width=95mm]{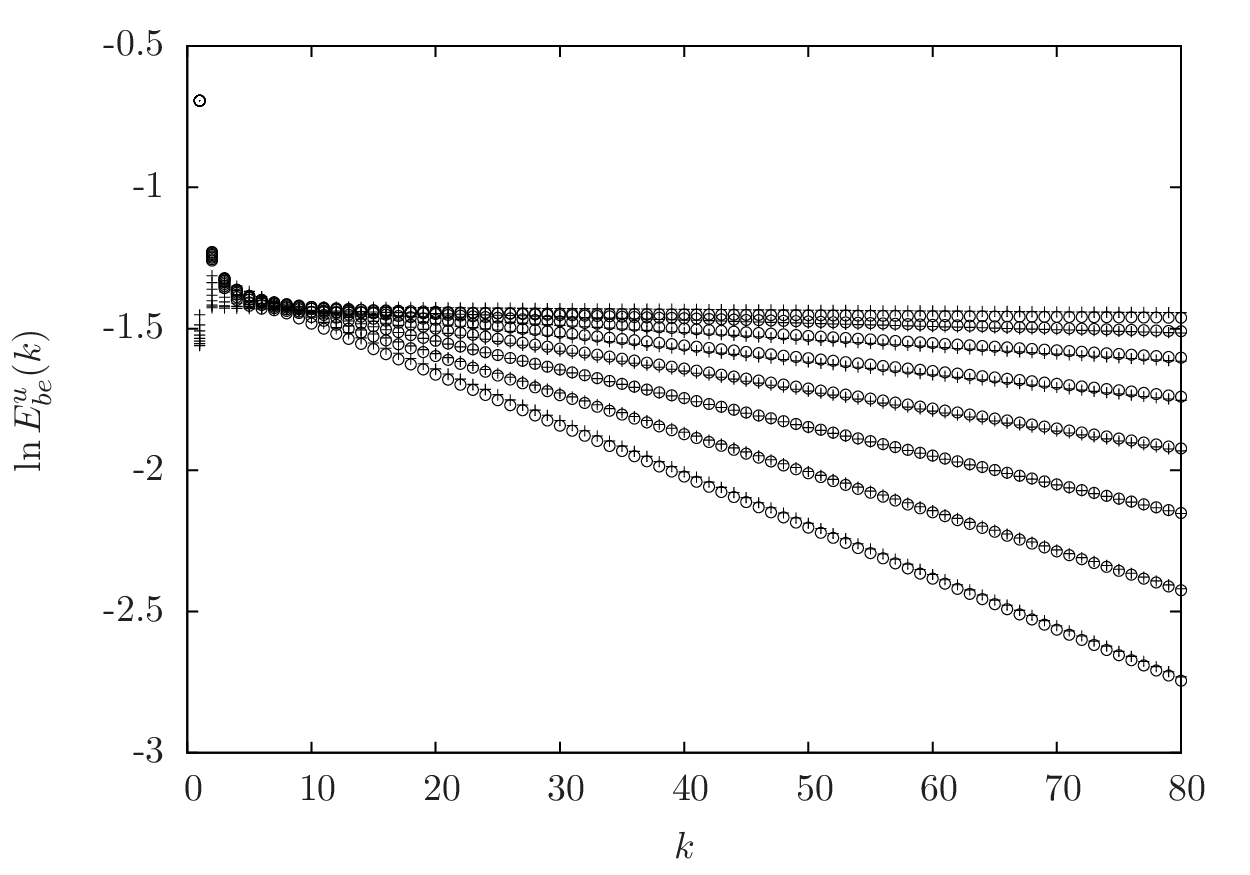}
\caption{Decay of the kinetic energy with $k$ for backward exponentially tapered
chains for $q = 0.01$ to $0.08$ from bottom to top, in steps of $0.01$. The circles correspond to the numerical data and the + symbols represent the BCA. \label{fig:E-k-exp-log}}
\end{figure}

We can similarly use the BCA to obtain the kinetic energy. The expression valid for all $k$
may be written as
\begin{equation}
  E_{be}^u(k) = \left( \frac{1}{2} \right)^{3-2k}
\left[ \left( 1-q \right)^{3/2} + \left( 1-q \right)^{-3/2}\right]^{2\left( 1-k \right)}.
  \label{eq:energy-binary-exp}
\end{equation}
Thus, for large $k$ we have
\begin{equation}
{\mathcal E}_{be}^u \equiv \frac{\ln[E_{be}^u(k)]}{k} \sim -2  \ln \frac{1}{2} \left[ \left( 1-q \right)^{3/2} + \left( 1-q \right)^{-3/2}\right].
  \label{eq:energy-binary-exp-asympt}
\end{equation}
We find that the BCA correctly predicts the exponential decay of the kinetic energy,
as shown in Fig.~\ref{fig:E-k-exp-log}, albeit not as accurately as the momentum prediction.
In Table~\ref{tab:coef-energy} we show the numerical comparison between the theory and the numerical
integration of the equations of motion.  The agreement is not as good as it is for
the momentum because the kinetic energy depends on the square of the velocity, and
the prediction of the BCA for the velocity is not accurate; hence
the error in the velocity is amplified.

\begin{table}[ht]
\centering
\begin{tabular}{|c|c|c|c|c|}
\hline
    & \multicolumn{2}{|c|}{Backward} & \multicolumn{2}{|c|}{Forward}  \\ \hline

$q$ & BCA & Numerical & BCA & Numerical \\ \hline \hline
0.01 &  	-0.00023 &  	-0.00033 &  -0.00022 &	-0.00042\\ \hline
0.02 &  	-0.00092 &  	-0.00118 & -0.000882 &	-0.00160\\ \hline
0.03 &  	-0.00209 &  	-0.00259 & -0.001965 &	-0.00375\\ \hline
0.04 &  	-0.00375  &  	-0.00455 & -0.003459 &	-0.00705\\ \hline
0.05 &  	-0.00591  &  	-0.00708 & -0.005351 &	-0.01164\\ \hline
0.06 &  	-0.00860  &  	-0.01017 & -0.007630 &	-0.01768\\ \hline
0.07 &  	-0.01183   &  	-0.01382 & -- & -- \\ \hline
0.08 &  	-0.01560   &  	-0.01805 & -- & -- \\ \hline
\end{tabular}
\caption{Comparison of decrease of the kinetic energy as predicted by the
BCA theory and the results obtained from the numerical integration of the equations of motion.}
\label{tab:coef-energy}
\end{table}

It should be stressed that in spite of the velocity decrease (as a power law
for the linearly tapered chain, exponentially for the exponentially tapered chain)
the momentum increases for both backward tapered chains as a function of $k$.
The energy, on the other hand, may decrease (exponential chain) or saturate
(linear chain). With respect to the tapering parameters, we notice that the larger
the values of $q$ or $S$, the larger is the momentum, but the smaller are the
energy and velocity.  Note that the momentum may be the important quantity
to focus on when studying a shock absorber.  Imagine our chain ending at a wall.
If the interaction of the last particle of the chain with the wall is perfectly elastic, then the
momentum traveling down the chain would be transferred to the wall.  In this case the
backward tapered chains would not be good shock absorbers; on the contrary, they magnify the shock.
These statements would need to be appropriately modified if the interaction of the last particle with the wall is inelastic.

\section{Forward tapered chains}
\label{for}
For forward tapered chains, the granules progressively decrease in size.  In linearly tapered chains the radii decrease arithmetically,
\begin{equation}
R_k^{fl}=1-S(k-1)
\end{equation}
where, as before, $S$ is the tapering parameter. This tapering imposes a limit on the number
of granules that may compose the chain to avoid the absurd situation of granules with negative
radii. Thus, the number $N$ of
granules in the chain must obey the restriction
\begin{equation}
  N < 1 + \frac{1}{S}.
  \label{eq:restriction}
\end{equation}
For exponentially tapered chains the radii decrease geometrically,
\begin{equation}
R_k^{fe} = \frac{1}{(1+q)^{k-1}}.
\end{equation}
For exponentially tapered chains a restriction on chain length is only imposed
for practical reasons (not to have granules that are too small to deal with experimentally,
and to avoid numerical errors in the analysis).  
 
\subsection{Linearly tapered chains}
\begin{figure}[ht]
\centering
\includegraphics[width=95mm]{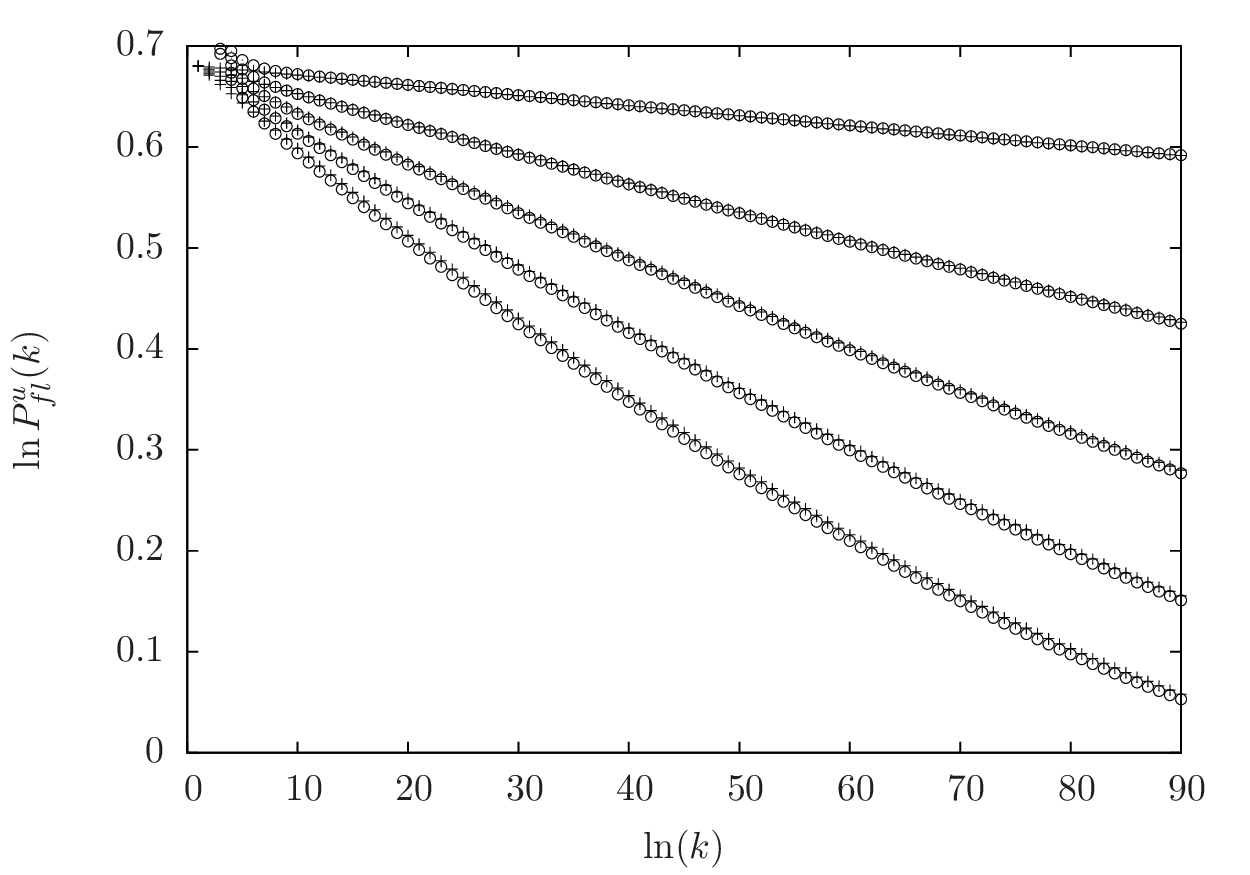}
\caption{Momentum as a function of $k$ for forward linearly
tapered chains is shown for $S = 0.001$ to $0.009$ from bottom to top, in
steps of $0.002$. The circles correspond to the numerical data and the + symbols represent the
BCA. \label{fig:p-k-lin-forward-fit}}
\end{figure}

\begin{figure}[ht]
\centering
\includegraphics[width=95mm]{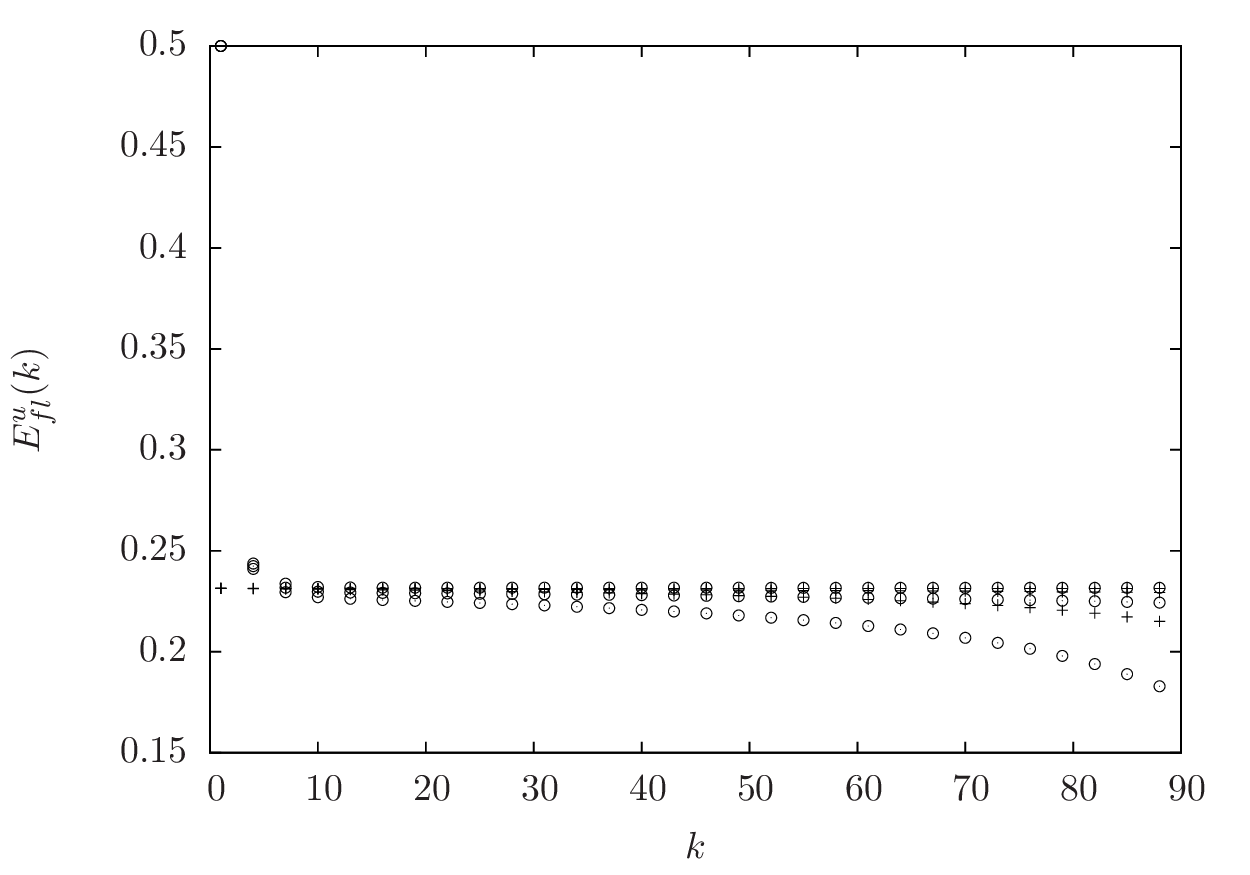}
\caption{Kinetic energy as a function of $k$ for forward linearly
tapered chains is shown for $S = 0.001$ to $0.009$ in steps of $0.004$ from top to bottom.The circles correspond to
the numerical data and the + symbols represent the BCA. \label{fig:E-k-lin-forward}}
\end{figure}

In this case, in the limit $k\ll 1 + 1/S$, the amplitude of the pulse velocity
in the BCA goes as~\cite{our-tapered}
\begin{equation}
  v_{fl}^u(k) \simeq \left( 1 - \frac{k}{1+1/S} \right)^{-3/2},
  \label{eq:vk-lin-forw}
\end{equation}
and the mass is
\begin{equation}
  m_{fl} (k)= \left[ 1 - S\left( k-1 \right)  \right]^{3} = \left( 1 + S \right)^3 \left[ 1 - \frac{k}{1+1/S} \right]^3.
  \label{eq:mk-lin-forw}
\end{equation}
Hence the momentum and kinetic energy are predicted to be, respectively,
\begin{eqnarray}
  P_{fl}^u(k) &\sim& S^3 \left( 1 + 1/S \right)^3 \left[ 1 - \frac{k}{1+1/S} \right]^{3/2} \nonumber\\
  &\simeq& S^3 \left( 1 + 1/S \right)^3 \left[ 1 -\frac{3}{2} \frac{k}{1+1/S} \right], \\
  E_{fl}^u(k) &\sim& 1.
  \label{eq:mom-ene-lin-forw}
\end{eqnarray}
Here, as in the case of backward tapered chains, we present 
the asymptotic formulas for the velocity, momentum and energy, but one 
should keep in mind that closed formulas can be obtained~\cite{our-tapered}.
In Fig.~\ref{fig:p-k-lin-forward-fit} we show that, for $k$ and $S$ not too large,
the momentum indeed decreases linearly with increasing $k$. Furthermore, for
sufficiently small $S$ the slope of the line is in fair agreement with the
BCA predictions (see Table~\ref{tab:momentum}).
The energy, shown in Fig.~\ref{fig:E-k-lin-forward}, is approximately constant
for small $S$ and $k$ not too large.

\begin{table}[ht]
\centering
\large
\begin{tabular}{|c|c|c|}
\hline
$q$ & Binary & Numerical \\ \hline \hline
 0.001& -0.0015&	-0.0010\\ \hline
 0.002& -0.0030&	-0.0020\\ \hline
 0.003& -0.0045&	-0.0030 \\ \hline
 0.004& -0.0060&	-0.0039 \\ \hline
 0.005& -0.0076&	-0.0047\\ \hline
 0.006& -0.0091&	-0.0056 \\ \hline
 0.007& -0.0106&	-0.0064 \\ \hline
 0.008& -0.0122&	-0.0071\\ \hline
 0.009& -0.0137&	-0.0078\\ \hline
 0.01 & -0.0153&	-0.0085\\ \hline
\end{tabular}
\caption{Coefficient of the linear decay of the momentum for forward exponentially tapered chains.}
\label{tab:momentum}
\end{table}
\subsection{Exponentially tapered chains}
\begin{figure}[ht]
\centering
\includegraphics[width=95mm]{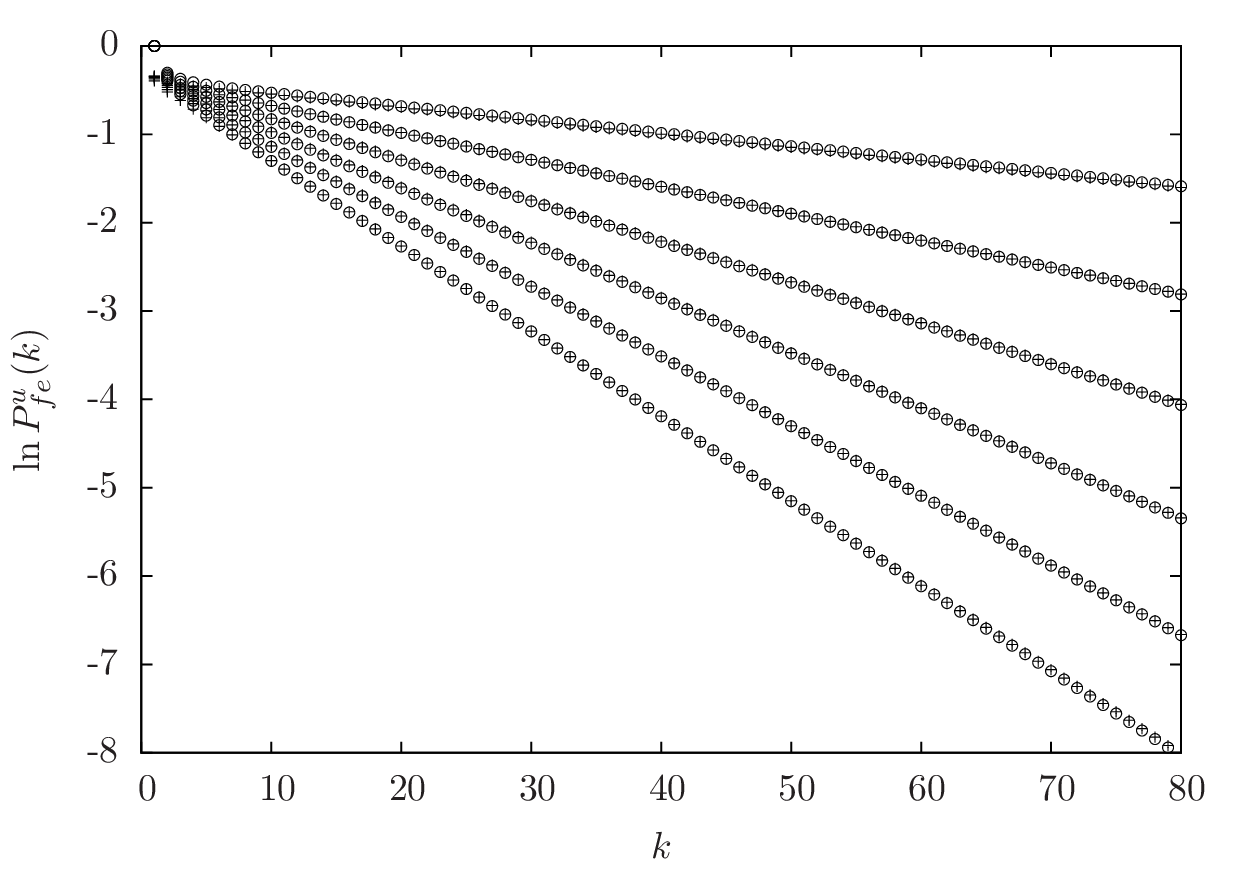}
\caption{Momentum as a function of $k$ for forward exponentially tapered
chains for $q = 0.01$ to $0.06$ from top to bottom, in steps of $0.01$.  The circles correspond to the numerical data and the + symbols represent the BCA.\label{fig:p-k-logscale-exp-forward}}
\end{figure}

For forward exponentially tapered chains we simply map $q \rightarrow -q$ in the
expressions for the backward tapered chains.
Hence, the momentum in the BCA is given by
\begin{equation}
  P_{fe}^u(k) = \left( 1 + \frac{3}{2} q + \frac{3}{2} q^2 + \frac{1}{2} q^3  \right)^{1-k}.
  \label{eq:momentum-exp-bin-forward}
\end{equation}
Thus, for forward tapered chains the BCA predicts an exponential
decay of the momentum as opposed to the growth in the backward chains. This is
observed in the numerical integration of the equations of motion, as can be seen
in Fig.~\ref{fig:p-k-logscale-exp-forward}. The rate of decay is in good agreement
with the BCA prediction (see Table~\ref{tab:coef-momentum}).

\begin{figure}[ht]
\centering
\includegraphics[width=95mm]{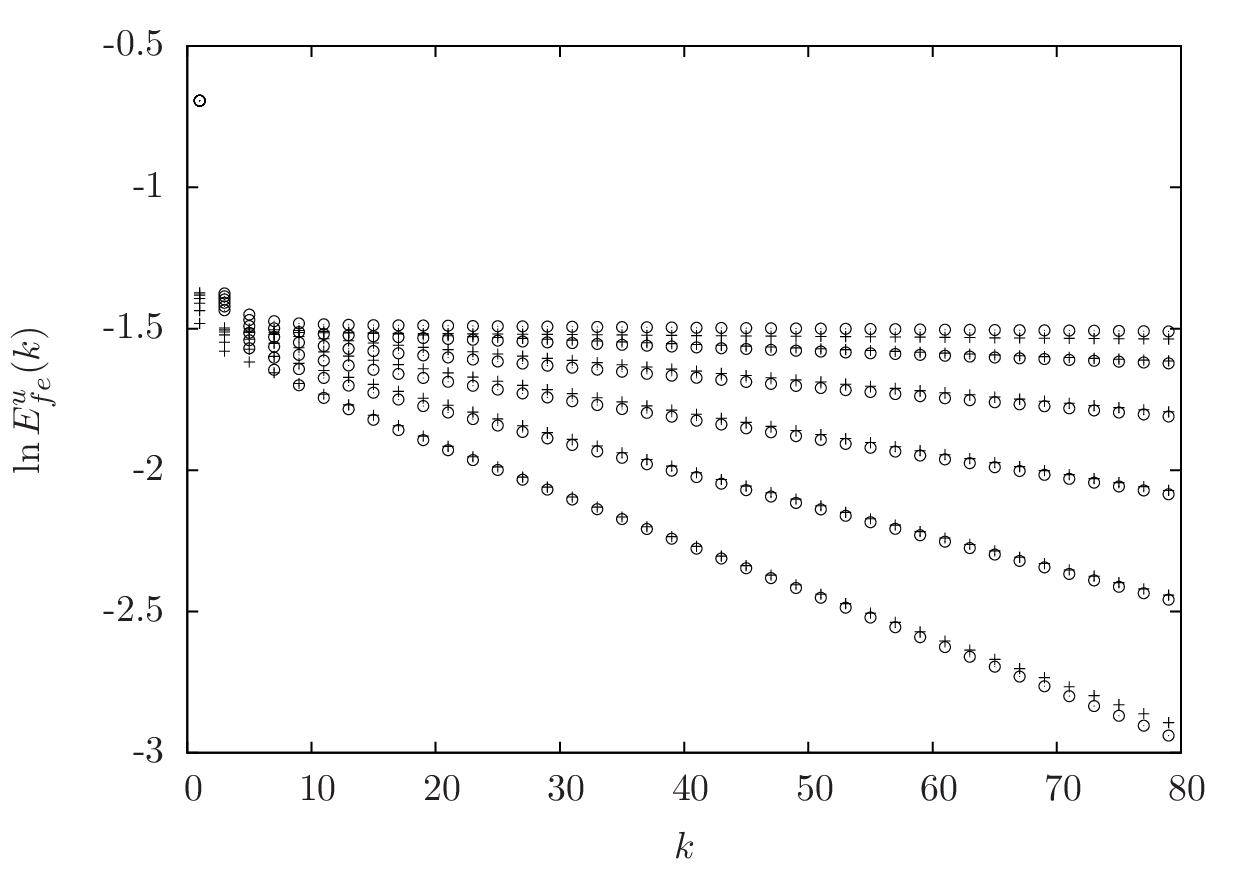}
\caption{Decay of kinetic energy pulse amplitude with $k$ for forward
exponentially tapered chains with $q = 0.01$ to $0.06$ from top to bottom, in steps of $0.01$. The circles correspond to the numerical data and the + symbols represent the BCA.
\label{fig:E-k-exp-log-forward}}
\end{figure}

For the kinetic energy, on the other hand, the BCA for large $k$ leads to
\begin{equation}
  \ln(E_{fe}^u)(k) \sim -2 k \ln \frac{1}{2} \left[ \left( 1+q \right)^{3/2} + \left( 1+q \right)^{-3/2}\right].
  \label{eq:energy-binary-exp-asympt-forward}
\end{equation}
Thus, as for the backward tapered chains, the BCA predicts an exponential decay of
the kinetic energy, which is corroborated by our data (see Fig.~\ref{fig:E-k-exp-log-forward}).
Once more, the agreement is not as good as for the momentum because the kinetic
energy depends on the square of the velocity, and hence the error in the velocity is
magnified (see the third and fourth columns of Table~\ref{tab:coef-energy}).

We end this section by assessing the behavior of the kinetic energy and momentum for the
different tapering configurations. First we notice that the kinetic energy is almost
constant (after a short transient) for both linearly tapered chains, while it decays
exponentially for both of the exponentially tapered chains. Thus, if one wishes to
distribute the energy more uniformly or minimize the energy first arriving at the end of
the chain, one should choose an exponentially tapered (in either direction) chain.
The momentum, on the other hand, increases for backward tapered chains, but
decreases for forward tapered chains. Therefore, if one wants to build a shock
absorber, one should choose a forward tapered chain. Moreover, exponentially
tapered chains are better candidates for this purpose since they lead to an
exponential decrease of the momentum, in contrast with the linear decay in the
linearly tapered chains.

\section{Modified Binary Collision Approximation}
\label{newresults}

Although the binary collision approximation correctly captures the trends of increase or
decrease of the kinetic energy and momentum, the actual values of these quantities are
not always in sufficient agreement with the results obtained from the numerical integration
of the equations of motion to be of predictive value. In this section we propose a
numerical rescaling of the velocity predicted by the BCA which leads to quantitatively
satisfactory predictors for the kinetic energy and momentum. Our error estimates are
in part, but only in part, numerical in the sense that the dependence of the
corrections on the tapering parameters is given analytically.  We thus propose
corrections that can then be used for all appropriate
situations once the tapering-parameter-independent coefficients and exponents
have been determined numerically once and for all.

As noted above, the BCA successfully captures the qualitative
behavior of the maximum velocity of the grains as the pulse moves along the chain. 
In proposing a correction that will improve the quantitative agreement with numerical
simulations, we have therefore assumed that the functional dependence (exponential or power law)
of the velocity $v_k$ on the grain position $k$ is correct as given by the theory, and that
the quantitative differences arise because the constants that characterize these functional
dependences require adjustment. We thus allow small adjustments of these constants,
explicitly, for example, of the exponent of $k$ and the proportionality constant $a_{bl}$  
in Eq.~(\ref{eq:vb-linear-back}) below.  We modify $a_{bl}^u$ to a value $a_{bl}$ presumed to lie close to
$a_{bl}^u$; similarly, the exponent of $k$ is perturbed by the small constant
$\delta_{bl}$, cf. Eq.~(\ref{eq:vs-linear-back}). We expect that similar corrections should
work for other tapering choices once the BCA has been used to establish the
$k$-dependence of the velocity.

In this section we will continue to denote all original BCA quantities by a superscript $u$ for unmodified
and the corrected velocities by quantities without a superscript.

\subsection{Backward tapered chains}
\subsubsection{Linearly backward tapered chains}

For linearly backward tapered chains (indicated by a subscript $bl$), according to the BCA in the large $k$ limit
the velocity of the granules increases as
\begin{equation}
v_{bl}^u(k) \simeq a_{bl}^u k^{-3/2}
  \label{eq:vb-linear-back}
\end{equation}
[cf. Eq.~(\ref{vltc})], where the constant $a_{bl}^u$ may depend on the tapering parameter $S$. We assume a
modified form
\begin{equation}
 v_{bl}(k) \simeq a_{bl} k^{-3/2 + \delta_{bl}},
\label{eq:vs-linear-back}
\end{equation}
where the constants $a_{bl}$ and $\delta_{bl}$ may again depend on the tapering parameter.
If this is correct, then the ratio of these velocities should be
\begin{equation}
 \frac{v_{bl}(k)}{v_{bl}^u(k)} \simeq \frac{a_{bl}}{a_{bl}^u} k^{\delta_{bl}}.
 \label{eq:ratio-v-linear-back}
\end{equation}
For small $\delta_{bl}$ as measured by the condition $\delta_{bl} \ln k \ll 1$, we can expand
$k^{\delta_{bl}} = e^{\delta_{bl} \ln k} \simeq 1 + \delta_{bl} \ln k.$ Consequently,
except for a logarithmic correction, we expect the ratio of the velocities to
be independent of granule number $k$, $v_{bl}(k)/v_{bl}^u(k)\simeq a_{bl}/a_{bl}^u$.

\begin{figure}[ht]
\centering
\includegraphics[width=95mm]{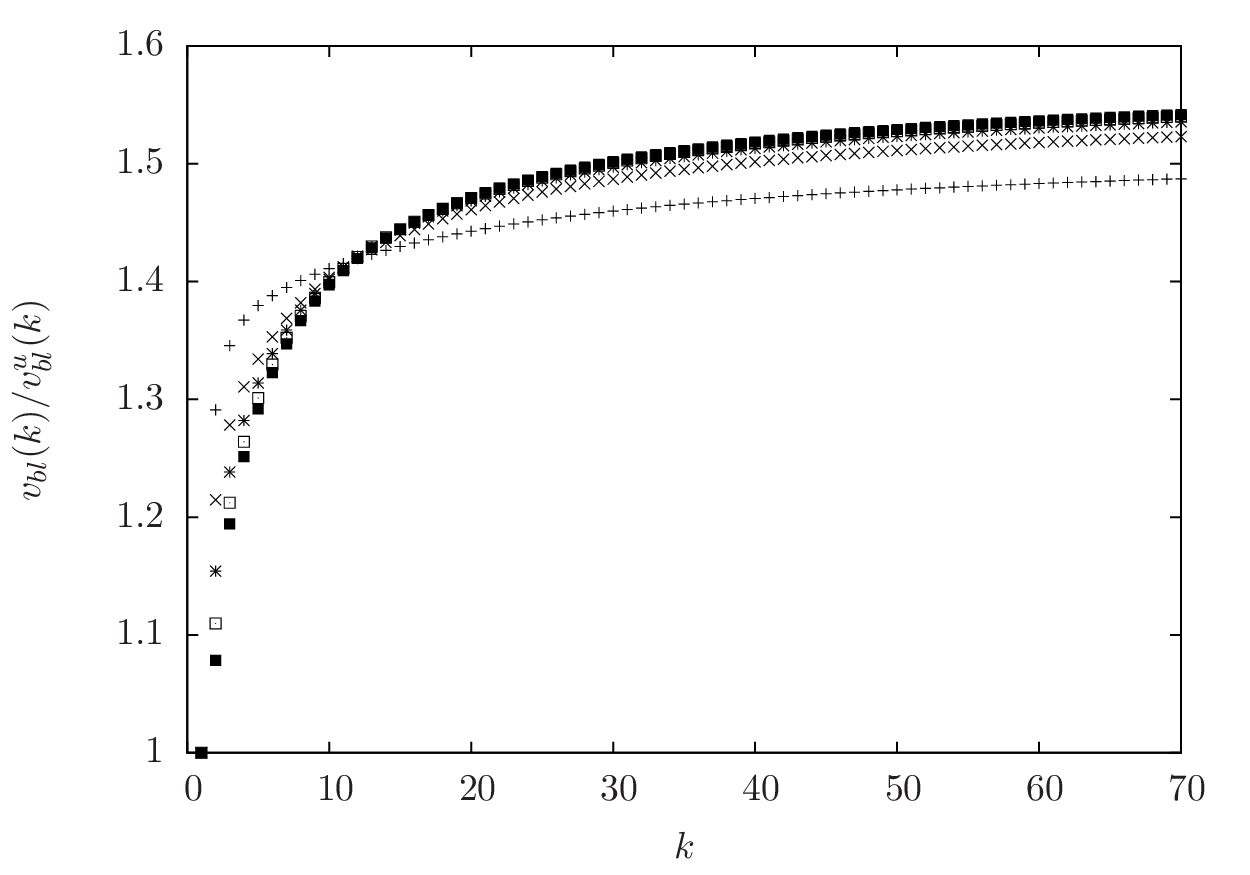}
\caption{Ratios of the velocity from the modified BCA and from
numerical integration data for backward linearly tapered chains for
$S = 0.1$ to $0.9$ from bottom to top at the right end of the figure,
in steps of $0.2$.\label{fig:ratio-lin}}
\end{figure}

\begin{figure}[ht]
\centering
\includegraphics[width=95mm]{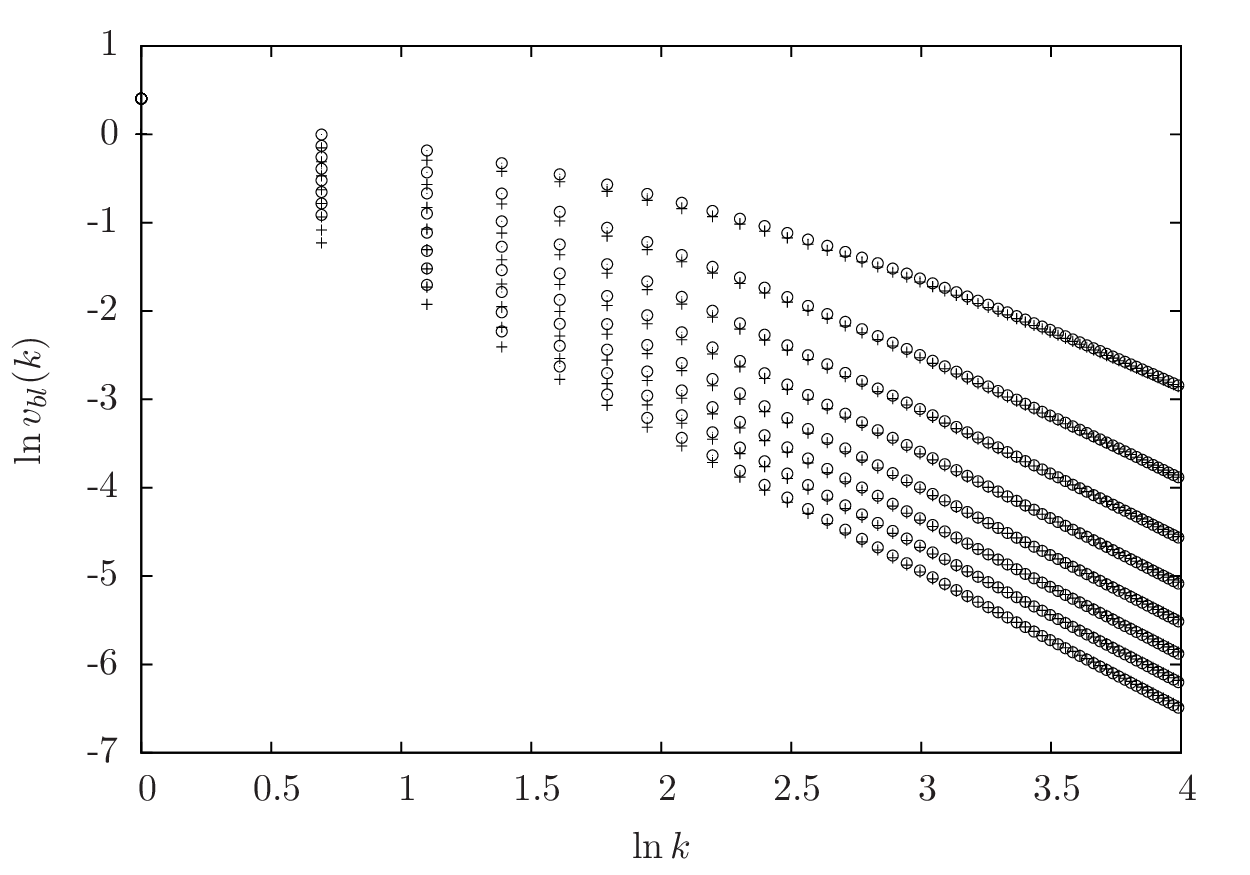}
\caption{Comparison between the modified BCA velocities ($+$ symbols) and
those obtained from the numerical integration of the equations of motion (circles) in
backward linearly tapered chains for
$S = 0.1$ to $0.9$ from top to bottom, in steps of $0.2$.\label{fig:v-k-lin}}
\end{figure}

In Fig.~\ref{fig:ratio-lin} we show that this ratio is indeed almost constant, not only
essentially independent of $k$ at sufficiently large granule number but also essentially
independent of the tapering parameter $S$.  As a working value we
take the ratio to be around $2/3$.  
With this rescaling, we see excellent agreement of the velocity
(Fig.~\ref{fig:v-k-lin}), momentum 
(Fig.~\ref{fig:p-k-logscale-lin}) and kinetic energy (Fig.~\ref{fig:E-k-asympt-lin}) with the results of the numerical integration of the equations
of motion. {\em This rescaling can henceforth be used predictively} for all linearly backward tapered chains: 
\begin{equation}
v_{bl} \simeq \frac{2}{3} k^{-3/2}.
\end{equation}

\subsection{Exponentially backward tapered chains}

\begin{figure}[ht]
\centering
\includegraphics[width=95mm]{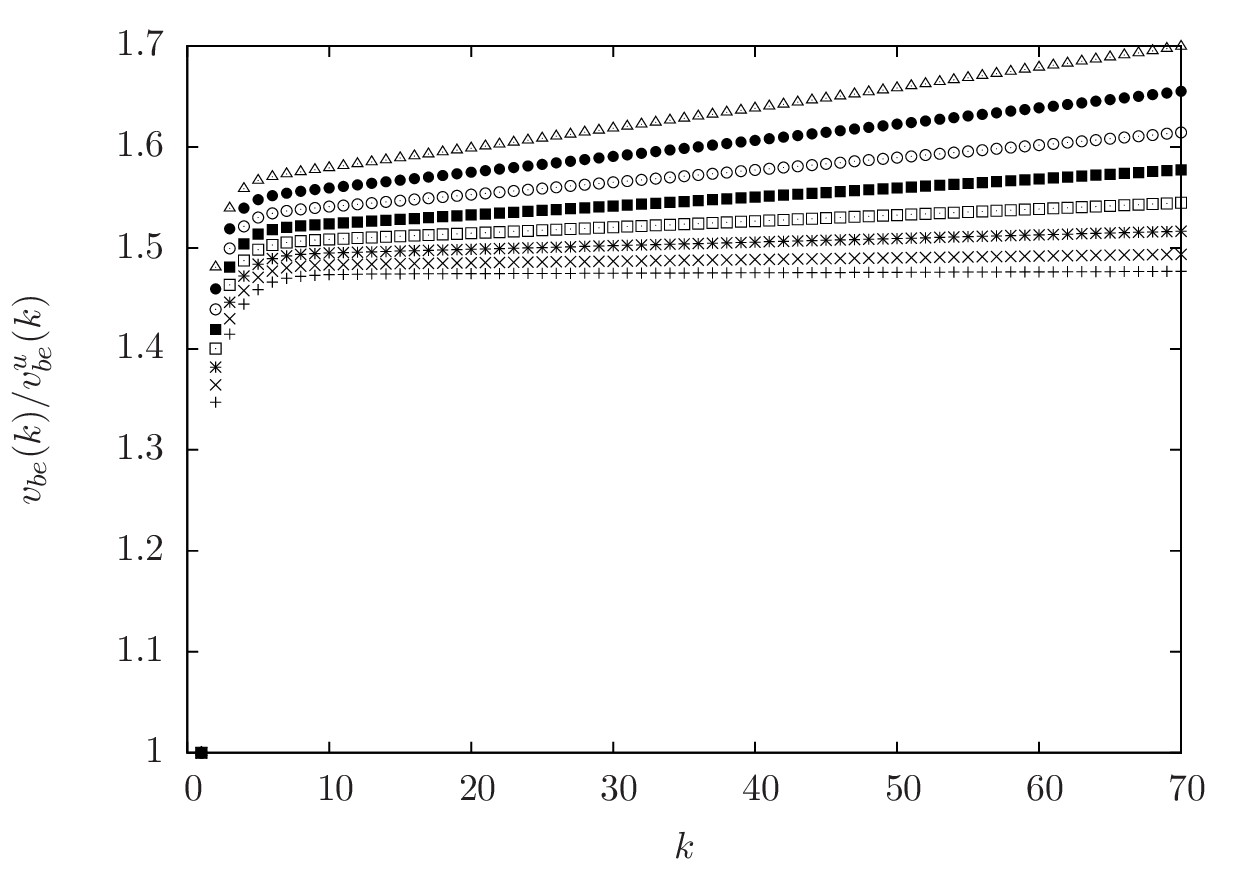}
\caption{Ratio of the modified BCA velocity over the velocity from the numerical integration
in backward exponentially tapered chains for
$q = 0.01$ to $0.08$ from bottom to top, in steps of
$0.01$.\label{fig:ratio-exp}}
\end{figure}

For the exponentially tapered chain, the velocity increases exponentially.
We write the BCA and modified BCA velocities as
\begin{eqnarray}
  v_{be}^u(k) &=& a_{be}^u e^{\alpha_{be}^u k},\\
  v_{be}(k) &=& a_{be} e^{\alpha_{be} k},
  \label{eq:vb-vs-exp-forward}
\end{eqnarray}
where $a_{be}^u$ and $\alpha_{be}^u$ are analytically known $k$-independent constants that depend on
the tapering parameter $q$, cf. Eq.~(\ref{vetc}), and $a_{be}$ and $\alpha_{be}$ are also assumed to be
independent of $k$.  We assume $|\alpha_{be}-\alpha_{be}^u| k \ll 1$ (a stronger restriction than the logarithmic one of the previous subsection).
Hence, the ratio of the velocities is
\begin{equation}
  \frac{v_{be}(k)}{v_{be}^u(k)} = \frac{a_{be}}{a_{be}^u} e^{|\alpha_{be}-\alpha_{be}^u| k} \simeq c_{be} + d_{be} k,
  \label{eq:ratio-exp-back}
\end{equation}
where $c_{be}$ and $d_{be}$ are assumed to be $k$-independent. In other words, we expect a linear
increase of the ratio of the velocities with $k$ (the correction here is linear instead of
logarithmic), as is indeed ascertained in Fig.~\ref{fig:ratio-exp}. The coefficients
in Eq.~(\ref{eq:ratio-exp-back}) do depend on the tapering parameter $q$, as clearly seen in the
figure. Note that the slope with granule number in the figure is mild, indicating that
$d_{be}$ is small for all $q$.

For this approach to be predictive, it is necessary to parametrize the $q$ dependence of the
coefficients.  We find that the dependence of $c_{be}(q)$ and $d_{be}(q)$ on $q$ is well described by
the power laws
\begin{equation}
  c_{be}(q) = c_{be}^0 q^{x_{be}} \quad  \mathrm{and} \quad d_{be}(q) = d_{be}^0q^{y_{be}}.
\end{equation}
We can obtain $c_{be}^0$ and $d_{be}^0$ by a linear fit of
\begin{equation}
\log c_{be}(q) = \log c_{be}^0 +  x_{be} \log q,
\end{equation}
and similarly
\begin{equation}
\log d_{be}(q) = \log d_{be}^0 +  y_{be} \log q.
\end{equation}
We find the values
\begin{equation}
c_{be}^0=1.64, \quad x_{be} = 0.0259, \quad d_{be}^0=0.211, \quad y_{be}= 1.83.
\end{equation}
In terms of these parameters the modified BCA can be written as
\begin{equation}
  v_{be} = \frac{v_{be}^u}{1.64q^{0.0259} + 0.211q^{1.83}k}.
\label{eq:vb-corrected}
\end{equation}

\begin{figure}[ht]
\centering
\includegraphics[width=95mm]{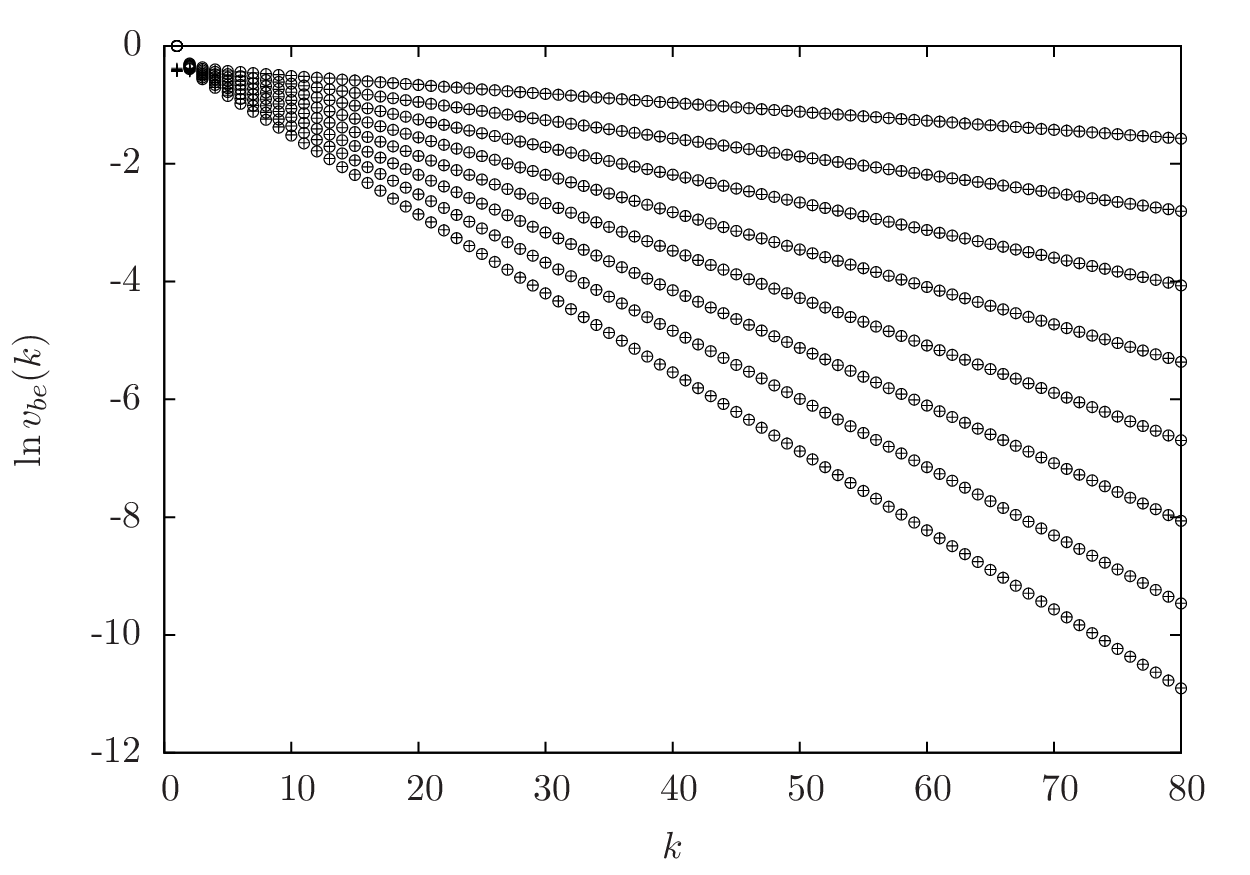}
\caption{Comparison between the velocities predicted by the modified BCA ($+$ symbols),
Eq. (\ref{eq:vb-corrected}), and numerical data (circles) in backward exponentially tapered chains for
$q = 0.01$ to $0.08$ from top to bottom, in steps of $0.01$.
\label{fig:v-k-exp}}
\end{figure}

The next figure shows that there is now excellent agreement between the results of the
modified BCA and those obtained by integrating the equations of motion.  Specifically, we direct
attention to Figs.~\ref{fig:v-k-exp} (velocity), \ref{fig:p-k-logscale-exp} (momentum)
and~\ref{fig:E-k-exp-log} (kinetic energy).

\subsection{Forward tapered chains}
\subsubsection{Linearly forward tapered chains}
\begin{figure}[ht]
\centering
\includegraphics[width=95mm]{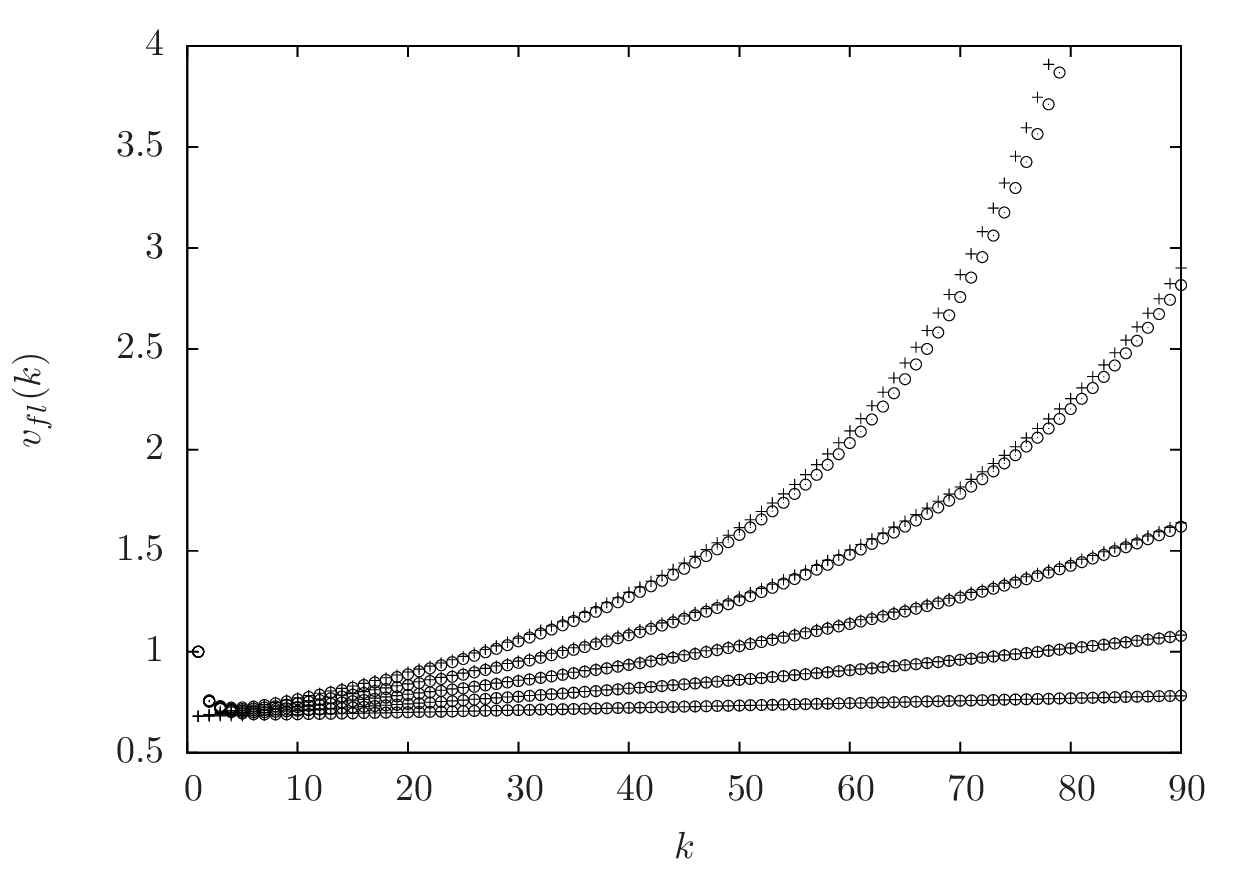}
\caption{Comparison between the velocities predicted by the modified BCA ($+$ symbols)
and numerical data (circles) in forward linearly tapered chains for $S = 0.001$ to $0.009$ from
bottom to top, in steps of $0.002$.\label{fig:v-k-lin-forward}}
\end{figure}

In this case, for $k \ll 1 + 1/S$ the BCA for the amplitude of the velocity is given
in Eq.~(\ref{eq:vk-lin-forw}).
Consequently, assuming that the simulation data can be described by
$v_{fl}^u(k) \sim (a_{fl}^u + b_{fl}^u k)^{-3/2}$ with constant $a_{fl}^u$ and $b_{fl}^u$, the leading term of the ratio $v_{fl}(k)/v_{fl}^u(k)$
is a constant, which we estimate as 0.680 (figure not shown). Using this factor to modify the
BCA result for the amplitude of the pulse velocity, we observe that the agreement
with the numerical results improves greatly for the velocity (Fig.~\ref{fig:v-k-lin-forward}),
the momentum (Fig.~\ref{fig:p-k-lin-forward-fit}), and the energy (Fig.~\ref{fig:E-k-lin-forward}) except for the larger values of
$S$, where slight deviations are noticeable. 

Melo and collaborators~\cite{melo06} have presented interesting experimental results for
forward linearly tapered chains (the authors claim that they use exponentially tapered
chains, but the radii provided in their paper are closer to those of linearly
tapered chains). They observed a linear decay of the momentum compatible with our
findings for linear chains (see Fig.~\ref{fig:p-k-lin-forward-fit}), showing
that our theory captures the experimental observations.

\subsubsection{Exponentially tapered chains}
\begin{figure}[ht]
\centering
\includegraphics[width=95mm]{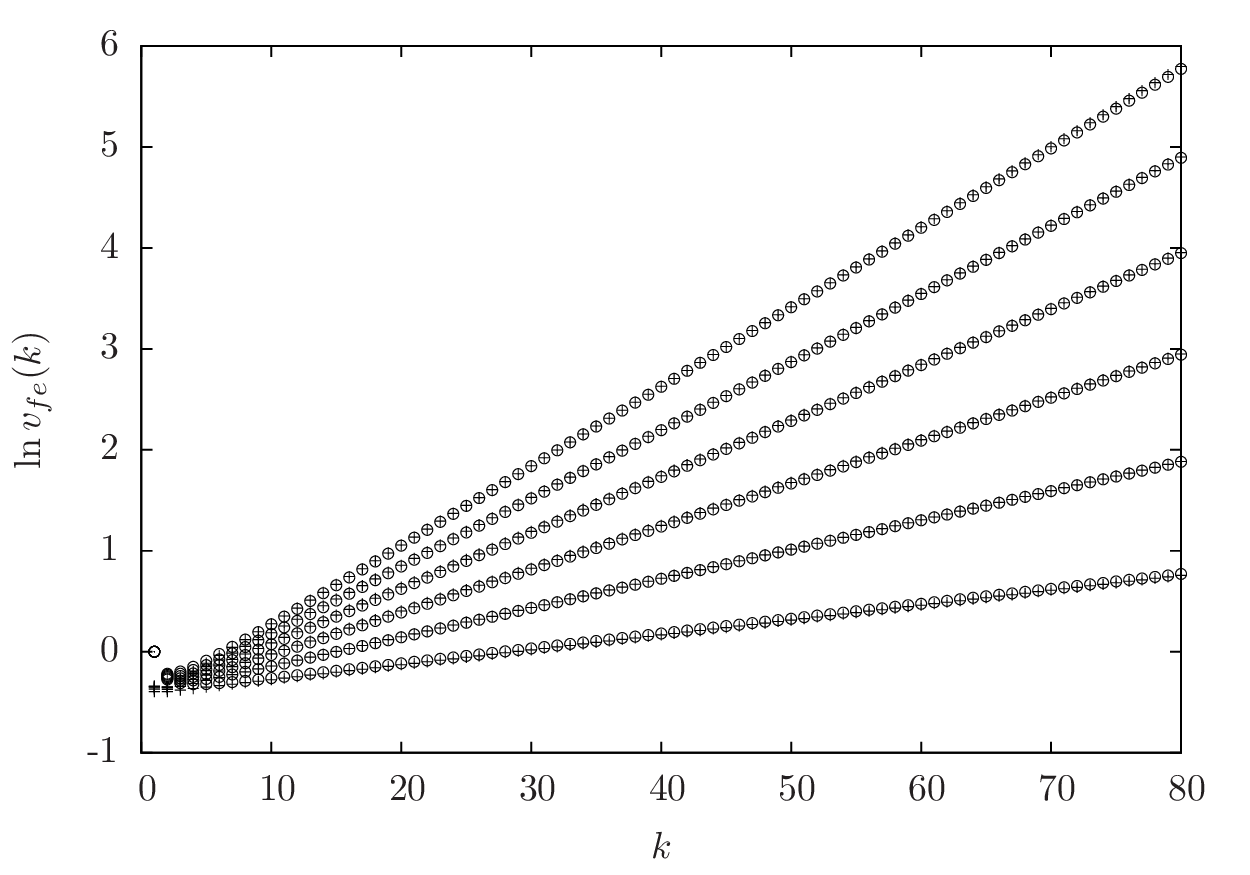}
\caption{Comparison between the velocity predicted by the modified BCA, Eq.~(\ref{eq:vb-corrected})
(+ symbols), and numerical integration results (circles) in forward exponentially
tapered chains for $q = 0.01$ to $0.06$ from bottom to top, in steps of
$0.01$.\label{fig:v-k-exp-forward}}
\end{figure}

For the forward exponentially tapered chain we have rescaled the velocity with the same
kind of function as in the backward case, cf. Eq.~(\ref{eq:vb-corrected}).
As noted earlier~\cite{our-tapered}, the forward tapering imposes a limitation on the
length of the chain. Therefore, we limited our study to tapering parameters $q$ no larger
than 0.06. We now find the values
\begin{equation}
c_{fe}^0=1.32, \quad x_{fe} = 0.0252, \quad d_{fe}^0=11.1, \quad y_{fe}=2.57.
\end{equation}
The scaled results are again in excellent agreement with the results of
the numerical integration of the equations of motion for the velocity
(Fig.~\ref{fig:v-k-exp-forward}). The momentum and kinetic energy as functions
of $k$ are shown in Fig.~\ref{fig:p-k-logscale-exp-forward} and Fig.~\ref{fig:E-k-exp-log-forward},
respectively.  The agreement is again seen to be excellent. 

At this point, it is worth noticing that this kind of tapering has been considered 
in~\cite{sen01,doney05,sen06} using a hard sphere approximation. However, the exceptional
 agreement between the numerical simulations and the theory achieved with our
formulation was not obtained within that approximation.

\section{Brief Summary}
\label{collection}

Recognizing that our analytic binary collision approximation, while highly accurate for the
prediction of a number of quantities that measure pulse propagation along tapered granular chains,
is not as accurate for others, we set out to improve upon the approximation.  In particular, we
noted that the propagation of momentum and of energy, two quantities that are of particular interest
in the design of shock absorbers, are among the latter.  Since tapered chains are often analyzed
with this particular application in mind, it is important to achieve analytic predictability for
these physical quantities.  

The binary collision approximation leads to analytic results and this is its strength, since it
overcomes the resource limitations posed by numerical and experimental searches of optimal
parameters.  It was therefore our hope to provide an analytic modification to the approximation that
could then be used for any tapering parameters. We succeeded to some extent.  While we found it
necessary to determine some of the coefficients in our corrective formulas numerically, these
quantities do not depend on the tapering parameters and can therefore be adopted once and for all
for predictive and optimization purposes. This claim must of course be understood within the
restrictions of the model: it is valid for Hertz potentials and for chains without precompression
(although the extension to mildly precompressed chains is straightforward~\cite{italo}). The theory
is valid for small values of the tapering parameters $S$ and $q$, which is not a serious restriction
since large values of these parameters would be experimentally difficult to implement.
Our corrected formulas work extremely well in almost all
cases, the weakest scenario being the kinetic energy in forward linearly tapered chains.  However,
we note that most of the literature that we are aware of, with the exception of the experiments of
Melo et al.~\cite{melo06}, focuses on exponentially tapered chains.  In all of these our new results
work exceedingly well.

We collect our results for convenience. We presented four kinds of tapered chains. The resulting
corrected pulse velocity formulas are as follows.
\begin{itemize}
\item
Backward linearly tapered chains
\begin{equation}
v_{bl}(k) = \frac{2}{3} k^{-3/2}
\end{equation}
\item
Backward exponentially tapered chains with tapering parameter $q$,
\begin{equation}
v_{be(k)}=\frac{\left\{(1/2)\left[1+(1-q)^{-3}\right]\right\}^{1-k}}{1.64q^{0.0259} + 0.211q^{1.83}k}
\end{equation}
\item
Forward linearly tapered chains
\begin{equation}
v_{fl}(k) = 0.680 \left(1-\frac{k}{1+1/S}\right)^{3/2}
\end{equation}
\item
Forward exponentially tapered chains with tapering parameter $q$,
\begin{equation}
v_{fe}(k)=\frac{\left\{(1/2)\left[1+(1+q)^{-3}\right]\right\}^{1-k}}{1.32q^{0.0252} + 11.1q^{2.57}k}
\end{equation}
\end{itemize}

We end by noting that
the momentum transferred by the pulse is attenuated in forward tapered chains,
making them good candidates for model studies of shock absorption. Moreover, the exponentially
forward tapered chains are stronger contenders since the momentum decrease is
exponential rather than linear (as in the linearly forward tapered chains).
Conversely, backward tapered chains act as momentum focusers such as might be desirable in sensing
tools, once again the exponential
tapered chains being the stronger contenders. We stress that these are of course model systems that
serve only to clarify how much more complex ``real" shock absorbers or momentum focusing devices
might be better understood. While one can imagine many other applications, at this point we make no
claims other than the utility of these systems as ones that help in the understanding of the
fundamental issues involved in energy propagation in granular matter.

\section*{Acknowledgments}
Acknowledgment is made to the Donors of the American Chemical Society Petroleum Research Fund for
partial support of this research (K.L.).  A.R. acknowledges support from Bionanotec-CAPES
and CNPq. L. P. M. acknowledges support by CAPES. The authors acknowledge helpful discussions with
A. H. Romero and U. Harbola.


\begin{thebibliography}{99}
\bibitem{note} The shape of the granules does not really matter as they are treated as point masses with an interaction determined by their radius of curvature and their mass.  We talk about spherical granules for convenience, as is done in a great deal of the relevant literature.
\end{thebibliography}
\end{document}